\documentclass[eqsecnum,aps,pra,twocolumn,groupedaddress,showpacs,showkeys]
{revtex4}
\usepackage{amssymb}
\usepackage{graphicx}
\usepackage{color}
\usepackage{epstopdf}
\graphicspath{{./figurearticle/}}

\begin{document}

\title{Measuring the influence of Casimir energy on superconducting
phase transitions: a cross-correlation data analysis}

\author{Annalisa Allocca$^{1}$ \thanks{Electronic address:
annalisa.allocca@gmail.com}, Giuseppe Bimonte$^{1,2}$ \thanks{Electronic
address: giuseppe.bimonte@na.infn.it}, Detlef Born$^{3}$,
Enrico Calloni$^{1,2}$ \thanks{Electronic address:
enrico.calloni@na.infn.it}, Giampiero Esposito$^{2}$ \thanks{Electronic
address:  giampiero.esposito@na.infn.it},
Uwe Huebner$^{3}$, Evgeni Il'ichev$^{3}$,
Luigi Rosa$^{1,2}$ \thanks{Electronic address: luigi.rosa@na.infn.it},
Francesco Tafuri$^{4}$}
\affiliation{
${\ }^{1}$Dipartimento di Scienze Fisiche, Universit\`{a} Federico II,\\
Complesso Universitario di Monte S. Angelo, Via Cintia, Edificio 6,
80126 Napoli, Italy \\
${\ }^{2}$ INFN, Sezione di Napoli, Complesso Universitario di
Monte S. Angelo, Via Cintia, Edificio 6, 80126 Napoli, Italy\\
${\ }^{3}$ Institute for Photonic Technology,
Postfach 10 02 39 - 07702 Jena, Germany\\ ${\ }^{4}$ Seconda
Universit\`a degli Studi di Napoli, Dipartimento di Ingegneria
dell'Informazione, via Roma 29, I-81031 Aversa (Ce), Italy}

\date{\today}

\begin{abstract}
The ALADIN experiment aims at observing  how the critical
magnetic field of a superconducting Aluminum film is modified, when it  constitutes one of the reflecting surfaces of a
Casimir cavity.  If successful, such an observation would reveal the influence of vacuum energy
on the superconducting phase transition.
In this paper a rigorous analysis of experimental data is reported,
the results are discussed and compared with theoretical predictions based on Lifshitz theory of dispersion forces, and the BCS
formula for the optical conductivity of superconductors.
The main novelty with respect to a previous data analysis  by some of the
authors, is the use of a cross-correlation method which is more rigorous
and leads to better estimates.
\end{abstract}

\maketitle

\section{Introduction}

The Casimir effect \cite{Casimir,Parsegian book} provides a striking  manifestation
of vacuum quantum fluctuations of the electromagnetic field in bounded geometries,
and it represents  a rare example of a purely quantum phenomenon that can be tested at the mesoscopic scale.
The Casimir effect has received  much
attention  in the past decades, thanks to  a wave of new
experiments which made it possible to measure the Casimir force with unprecedented precision.
For a recent review of these experiments and a critical survey of the numerous theoretical investigations
on the Casimir effect in   materials, we address the reader to the recent monograph \cite{bordag}.

Despite the  impressive theoretical and experimental advances made in the past twenty years,   the Casimir effect still
faces important unsolved questions at the fundamental level, in particular
the problem of reconciling the vacuum
energy density and its interaction with the gravitational field, known as
the cosmological constant problem \cite{weinberg_art, feynman}. No
experimental verification that vacuum fluctuations  gravitate according
to the equivalence principle has been obtained  so far, even though there
are theoretical expectations that this  should be the case \cite{brown, sciama, USA1,
USA2, USA3, USA4, USA5, USA6, USA7}. Relying upon
these considerations, some of us  studied the effect of a gravitational
field on a rigid Casimir cavity, by computing the net force acting on it:
interestingly,  it was found that a Casimir apparatus,
when subject to the weak gravitational field of the Earth, should
experience a tiny push in the upwards direction
\cite{vacuum_fluctuation_force,gravitational_effects,
energy-momentum_tensor,relativistic_mechanics}.
In \cite{vacuum_fluctuation_force} it  was  argued that an experimental verification of this effect is extremely hard, if not impossible,
under static conditions,  by virtue of the extreme smallness of the expected force. A better possibility would be to carry out the
measurement dynamically, i.e. by modulating the Casimir energy stored in a rigid
cavity in a known way. Such a modulation of the Casimir energy could  be achieved by  altering periodically the reflectivity of the plates.
Recently, a significant modulation of the Casimir force between  a highly doped semiconducting membrane and a gold plate has been demonstrated experimentally \cite{Umar}, by shining periodic laser pulses  on the membrane which determine a large change in the charge carrier density of the membrane.  Despite being interesting for Casimir studies, this result is not suitable for ``weighting" aims:  the energy supplied to the system to induce the change in carrier density is many orders of magnitude  larger than the variation of Casimir energy.  This would make it extremely difficult to  observe the tiny fraction of mass change due to the Casimir contribution. On the contrary, in  our scheme based on the superconducting phase transition, the total change of energy is of the same order of magnitude  as the change in the Casimir energy and  therefore its contribution   might in principle be  observed.

This is the framework of the ALADIN experiment, whose aim
is to observe the variation of the Casimir energy stored in a    superconducting Casimir cavity,
constituted by a thin superconducting film  separated by a thin oxide layer from a thick gold substrate,
across the superconducting phase transition. The scheme of detection is based
on a measurement of the critical magnetic field that destroys the superconductivity of the film,
whose magnitude is expected to be influenced by the Casimir energy. If successful, the experiment
would thus reveal the influence of vacuum energy on a phase transition.
Another distinctive feature of  our setup is that  we use  rigid cavities, that are obtained by deposition techniques,
a feature
which  might be useful   to investigate  experimentally the
dependence of the Casimir energy on the geometrical shape of the intervening bodies, an issue that  remains under
scientific debate also at a theoretical level \cite{bordag}.

The plan of the paper is a follows: in Sec. II we briefly describe
the experimental setup and the measurement method,
Sec. III is devoted to the analysis method based on cross-correlation.
Finally, the experimental results are discussed in Sec. IV.
Indeed, a preliminary analysis of the data reported
in this paper has already been performed some time ago, but without the necessary
rigor, and we reported on it in \cite{low_noise}. The analysis reported in this paper, based on a
procedure of cross-correlation, is more rigorous and the results are now better
estimated.

\section{ALADIN: experimental setup and expected effect}
\label{exp_setup}

Before we describe our experimental setup, it is useful to briefly recall the
principle  at the basis of our experiment,  that was described in detail in the works \cite{towards,variation}.   The starting
observation is that according to Lifshitz
theory \cite{bordag}, the Casimir energy is determined by the optical properties of the plates. Since
the optical properties  of a superconductor  are sharply different from those of a normal metal \cite{glover}, one is led to expect
that the Casimir free energy $F^{(C)}$ stored in a superconducting cavity should change across the superconducting transition. The change $\Delta F^{(C)}=F^{(C)}_n-F^{(C)}_s$ of
Casimir energy was estimated in \cite{towards,variation}  (see also \cite{bimonte,haakh}), on the basis of Lifshitz theory by using the BCS formula for the optical conductivity of superconductors,
 and it was found to be extremely small. This is not surprising of course, because the superconducting transition alters the optical properties of
 a metal only  in the microwave region,
which   constitutes  a small
window in the  wide frequency range that contributes to the Casimir energy. The latter typically extends up-to  a few times  characteristic cavity frequency $\omega_c=c/2 d$, with $d$  the plate separation,   which for typical submicron separations    belongs to the infrared region of the spectrum.
The smallness of the fractional change of Casimir energy across the superconducting transition makes it impossible to
observe the corresponding change in the Casimir force on the plates, with present day sensitivities in force
measurements.

The   Aladin experiment uses a detection scheme which does not involve at all a force measurement,  as it aims at observing how the variation    $\Delta F^{(C)}$ of Casimir energy
influences the critical magnetic field $H_c$
of a thin superconducting film which is part of a Casimir cavity. To see how this comes about, we recall \cite{tinkham}  that   the magnitude of the parallel critical field $H_{c}$
for a thick superconducting slab of volume $V$  can be determined by equating the magnetic work $V {H_{c}^{2}/ 8\pi}$ required to expel the magnetic field from the sample, with
the so-called condensation energy $\varepsilon_{\rm cond}(T)$ of the material, which represents the difference of Helmholtz free energies
between the normal and the superconducting phases:
\begin{equation}
V {H_{c}^{2}(T)\over 8\pi}=\varepsilon_{\rm cond}(T).
\end{equation}
When the film is one of the two plates of
a Casimir cavity, we have to augment the right-hand side
of the above Equation by the difference $\Delta F^{(C)}$ between the Casimir energies in the normal and in the superconducting phases:
\begin{equation}
V {H_{c}^{2}(T)\over 8\pi}=\varepsilon_{\rm cond}(T)
+\Delta F^{(C)}(T). \label{shiftcas}
\end{equation}
In writing this relation, we are tacitly assuming that the fluctuating electromagnetic field in the Casimir cavity does not
alter significantly the properties of the superconductor, and in particular its condensation energy. This may be considered as a plausible
assumption,   as far  as  $\Delta F^{(C)}(T) \ll \varepsilon_{\rm cond}(T)$.
According to Eq. (\ref{shiftcas}),  the variation $\Delta F^{(C)}(T)$ of Casimir energy  determines a change in the magnitude of the critical field $H_c$,
which for  $\Delta F^{(C)}(T) \ll \varepsilon_{\rm cond}(T)$ is estimated to be:
\begin{equation}
{\delta H_{c} \over H_{c}} \approx
{ \Delta F^{(C)}(T)\over
2 \,\varepsilon_{\rm cond}(T)}.\label{change}
\end{equation}
The key thing to notice is that the condensation energy of a thin superconducting film  can easily be orders of magnitudes smaller
than  typical  Casimir energies $F^{(C)}_n$, and therefore one may hope that even tiny fractional changes of Casimir energy $\Delta F^{(C)}(T)$
can determine observable shifts of the critical field.
For example,  for a Beryllium film, we estimated \cite{towards,variation} that a relative variation of $F^{(C)}$
of one part over $10^{8}$ might lead to a 5 percent variation of
critical magnetic field.

Detailed numerical computations   \cite{variation} show  that the magnitude of the effect increases for thin films, because they have a smaller condensation energy, and for
  small cavity widths $d$, because the change of Casimir energy becomes larger. It is important to remark that the relative shift of critical field  was found to be roughly proportional to the inverse of the
transition temperature $T_c$ of the superconducting material. The main reason why superconductors with low $T_c$ lead to a larger effect
is that the condensation energy
is empirically known to scale as $T_c^{2.6}$ \cite{lewis},  and this makes the denominator in the r.h.s. of Eq. (\ref{change})  decrease faster than the numerator as
$T_c$ decreases.
As a compromise between the possibility to perform preliminary tests,
easy change of structures, statistics and signal-to-noise ratio, we chose to
work around 1.5 K, using Al as superconducting material.
Besides having a low critical temperature, Al is a
material  that oxides easily,  and this makes it easier
to   grow  oxide layers of controllable thicknesses, well attached to the superconducting
film, that   constitute the dielectric medium of our metallic Casimir cavities.
The  configuration used in the experiment is a three-layer cavity, made of a thin
superconducting Al film ($5\div10$ nm), a thin dielectric layer of native
oxide ($Al_2O_3$)($5\div10$ nm), and a thick metallic layer of Au (100 nm). However, at present we are considering different configurations
in order to obtain a  larger signal-to-noise ratio in the expected effect.

\begin{figure}
\includegraphics[width=0.95\linewidth]{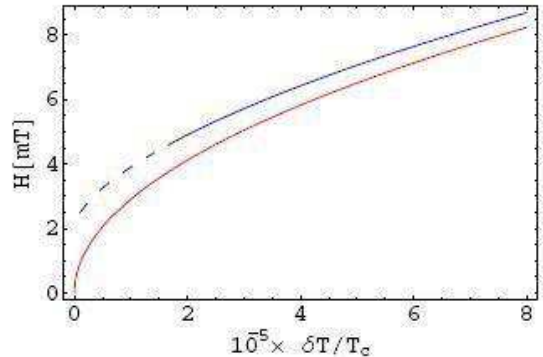}
\caption{Simulation of the expected signal for a bare thin Al film
of thickness $D=14$ nm (lower curve) and for a cavity consisting of a
similar Al film, covered by a 6 nm dielectric layer and a 100 nm Au
mirror (upper curve).}
\label{expected}
\end{figure}

In  our experiment, we  used a cryogenic system based on the
Heliox VL $^3He$ cryostat, inserted into a dewar equipped with magnetic
screening, which isolates the samples from external EM fields.
Since it is extremely difficult to keep the cryostat temperature perfectly constant,
we did not try to measure the  critical field  $H_c$ of the samples  as a function of the temperature. Rather,
we measure how the critical temperature  $T_c(H)$ of the samples {\it changes} as a function of the applied magnetic field $H$.
More precisely, for each value of the applied field $H$ we measure the relative {\it shift} $\delta t= (T_c(H)-T_{c \, 0})/T_{c \, 0}$ in the critical temperature $T_c$ of the sample,
with respect to the critical temperature $T_{c \, 0}$ of the same sample in zero field. As we shall explain below, by using the
method of {\it cross-correlations} the shift $\delta t$ can be measured much more accurately than the individual critical temperatures
in the applied and in the  null fields.
Our theory predicts (see Figure \ref{expected} and comments below) that the curve $ H (\delta t)$ for the bare film should lie below that for the film in the Casimir cavity.
Since this is a {\it differential} measurement,  we need a
very good sensitivity in temperature, of   order  a few $\mu K$ in the
case of Al. As described in \cite{low_noise},  the critical temperature is determined by
measuring the
resistance of a sample R(T) in a four-wire configuration around the phase
transition, for different external applied fields. Several
measuremements have been performed and different samples have been
tested, so as to find the best experimental conditions for a good
signal-to-noise ratio. For a good data analysis to be possible, it was important
to make sure  that the
transition curves did not change their profiles in time, or  after applying a
magnetic field. By employing these criteria,
the best samples have been selected. Data
reported in the following  were obtained from the samples showing the sharper
transition and the highest homogeneity among transition curves.
For a detailed description of sample preparation, cryogenic apparatus and
measurement scheme, see \cite{low_noise}.\\

The expected effect is shown in Fig. \ref{expected}.
Here the  critical magnetic field $H$ is plotted against the shift
$\delta t=1-t$ of the {\it reduced} critical temperature $t=T_c(H)/T_{c\,0}$
(with $T_{c\,0}$ zero-field transition temperature). We recall \cite{tinkham} that for a (bare) thin superconducting film, with a thickness $D$ much less than the penetration length $\lambda$, incomplete field expulsion leads to higher values of the parallel critical field $H$, as compared to bulk samples. For
$1-t\ll 1$ (as in our case), the thin-film critical magnetic field $H $ follows the simple law
\cite{tinkham}

\begin{equation}
H =H_0\sqrt{24}\frac{\lambda(0)}{D}\sqrt{1-t}\;,
\label{H_par}
\end{equation}
where $\lambda(0)$ is the penetration depth and $H_0$ is the bulk
  critical field, both determined for zero temperature. \\

For the in-cavity film we can divide the
expected signal curve (upper curve in Fig. \ref{expected}) into three temperature regions. For temperatures far
from the transition temperature of the film (region not shown in  Fig. \ref{expected})
the in-cavity curve coincides with the bare film curve, since the vacuum
energy contribution  becomes negligible compared to the condensation energy.
When $\delta t\approx 3\times 10^{-5}$,
the change of Casimir energy is small but no
longer negligible, and a perturbative approach is possible in $\delta H/H$:
for $H\approx 5\div 6$ mT the two curves are expected to differ by an
amount $\delta t\approx 6\times 10^{-6}$. Note that, since the critical
temperature of Al is 1.5 K, this corresponds to a
shift in temperature of order $10$
$\mu K$.  Finally, for lower temperatures the dependence of $\delta t$ on
$H $ for the in-cavity curve is expected to differ from the bare film
case, because the Casimir energy contribution is of the same order of
magnitude  as the condensation energy. In this region
we are not able to perform any perturbative
calculation, and there is no theoretical prediction for the
curve's trend (dashed line in  Fig. \ref{expected}).

\section{Analysis method based on cross-correlation}
\label{analisys}

\begin{figure}
\includegraphics[width=0.99\linewidth]{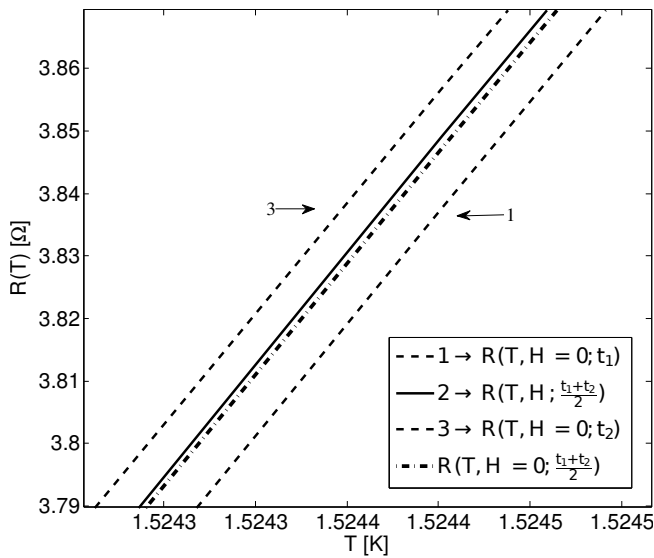}
\caption{Example of the measurement sequence. All measurements are
performed at equal time  intervals. Before and after each measurement with the
applied field, a measurement in the absence of field is performed. 
From the two measurements in absence of field, it is
possible to reconstruct the zero-field curve that would be measured
simultaneously to the transition curve in the presence of the magnetic field.
$\Delta T_c$ is the distance in temperature between the R(T) curve in the presence of the magnetic field
(continuous line), and the reconstructed R(T) curve in zero field (dashed-dotted line).}
\label{R(T)measur}
\end{figure}

As  was explained in the previous Section,
for each value of the applied magnetic field $H$ we need to determine accurately
the  fractional shift $\delta t$ of the critical temperature of the sample,  relative
to  its critical temperature $T_{c\,0}$ in zero field.
As described in Ref. \cite{low_noise}, the major
source of noise in our measurements is the electronic
noise at the read-out amplifier. This noise has a ``fast" component
(with time scale of one second) that determines a statistical error of a
few $\mu K$, and a slow thermal drift (linear in time, about 50 $\mu K$ per
hour) that produces a shift proportional to the time elapsed between two
measurements.
To correct for the thermal drift we arranged our measurements in series of triplets, as   is shown in Fig. \ref{R(T)measur}.   Each triplet consists of three measurements of the curve $R(T)$ that are  equally spaced in time, the first and the last one of which (left-most and right-most dashed curves in Fig. \ref{R(T)measur}) are performed in zero-field, while the intermediate one (continuous line in Fig. \ref{R(T)measur}) is done with the field applied.  Since the
slow thermal drift is linear in time, and since the two zero-field measurements are performed at equal time intervals before and after the applied field measurement, it is possible
to  ``reconstruct"  out of  them the position that the zero-field curve $R(T)$ would have occupied in the $R-T$ plane (dotted-dashed line in Fig. \ref{R(T)measur}), had it been measured at the same time as the applied field curve.  The relative shift  $\delta t$ in the critical temperatures is then determined by comparing the measured curve $R(T)$ in the applied field, with the reconstructed curve in zero field. Having explained the general scheme of the measurements, let us see how the method of cross-correlations permits to accurately determine
$\delta t$.

The key feature of the $R(T)$ curves  that allows to use the cross-correlation method to accurately determine the shifts $\delta  t$ is that, for each sample, their {\it shape} is
practically independent of both the intensity of the applied magnetic field, and   the time at which the measurements are  performed. In other words, the curves $R(T)$
that are taken at different times, or in different fields, appear to differ from each other just by some horizontal  translation $\Delta$  along the temperature axis. The cross-correlation method is ideally well-suited to determine the amount of this translation, independently of any  model for the shape of the curves. The idea is very simple, and consists in looking for the translation that {\it maximizes} the {\it overlap} between any two $R(T)$ curves. Let us see how this works out in detail. In reality, each  $R(T)$  curve consists of
a large number of data points more or less scattered in the $T-R$ plane, around some ideal transition curve. We consider only   data points that belong to the transition region, having a width of a
few mK, around the critical temperature. Typically, this region contains a few thousand points for each curve. At this point, we cover the $T-R$ plane by a rectangular grid whose axis are parallel to the $T$ and $R$ coordinate axis, and whose steps are $s_T$ and $s_R$ respectively. The points $p_{i,j}$ of the grid in the $T-R$ plane thus have coordinates $\{i\, s_T +a_t,j \,s_R + a_R\}$, where $\{a_T,a_R\}$ are the coordinates of the grid's origin.
 We let ${\cal H}_{i,j}$ be the number of data points that occupy the grid cell whose left-down corner coincides with the point $p(i,j)$ of the grid. Clearly, the number of points occupying the
 non-empty cells depends on the size of the cells, i.e. on the steps $s_T$ and $s_R$. We shall  discuss below the criterion we used to choose these steps. In practice for the chosen size of the steps,
 the occupied cells turn out to include about  five or six data points. Consider now any two transition curves $R^{(a)}(T)$ and $R^{(b)}(T)$ (not necessarily distinct), and let  ${\cal H}^{(a)}_{i,j}$ and ${\cal H}^{(b)}_{i,j}$ be the respective
 histograms. For any translation of $R^{(b)}(T)$ by $n$ steps along the $T$-axis, we define the   cross-correlation $\mathcal{C}^{(a,b)}[n]$,    to be the quantity:
\begin{equation}
\mathcal{C}^{(a,b)}[n]=\sum_{i,j}{\cal H}^{(a)}_{i,j}\,{\cal H}^{(b)}_{i-n,j} \;.
\label{crosscorrRT}
\end{equation}
The quantity ${C}^{(a)}[n] \equiv {C}^{(a,a)}[n]$ shall be denoted in what follows as the self-correlation of the curve $R^{(a)}(T)$.
Intuitively, $\mathcal{C}^{(a,b)}[n]$ measures how well  $R^{(a)}(T)$ and $R^{(b)}(T)$ overlap, after we translate  $R^{(b)}(T)$ by an amount $n \times  s_T$ along the $T$-axis.

\begin{figure}
\includegraphics[width=0.95\linewidth]{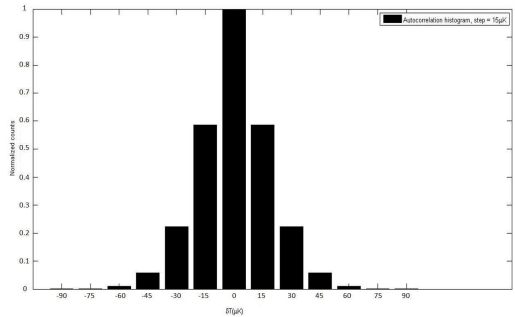}
\caption{Self-correlation of the $R(T)$ curve obtained with a
lattice step on the abscissa equal to $15 \mu$K.}
\label{selfone}
\end{figure}

\begin{figure}
\includegraphics[width=0.95\linewidth]{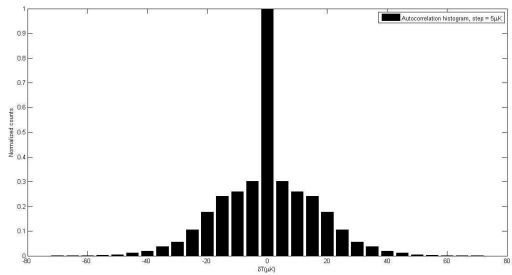}
\caption{Self-correlation of the $R(T)$ curve with lattice step
equal to $5 \mu$K.}
\label{selftwo}
\end{figure}

\begin{figure}
\includegraphics[width=0.95\linewidth]{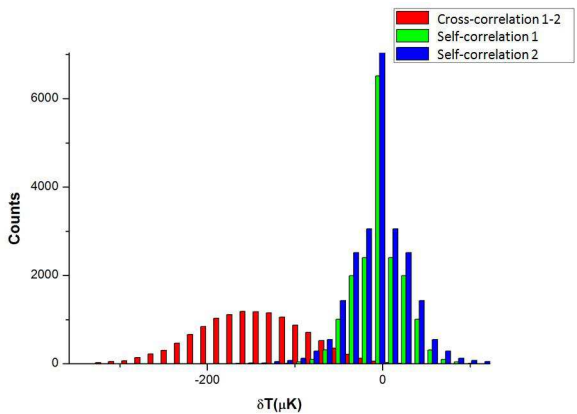}
\caption{Self-correlation of two $R(T)$ curves, green and blue,
and cross-correlation among the two.}
\label{cross}
\end{figure}

In Figs. \ref{selfone} and \ref{selftwo} we  plot the normalized self-correlation ${\cal C}[n]$ of one of our
$R(T)$ curves, for two different steps $s_T$. In Fig. \ref{cross} we plot
the self-correlations and the crossed correlations of two of our curves, one measured in zero field and the other
in a non-zero field. We verified that all correlations have a Gaussian shape around the central peak. More importantly,
we see from Fig. \ref{cross} that the maximum of the cross-correlation between the zero-field curve and the non-zero-field curve is of the same
order of magnitude as the maximum  of the self-correlations of the two curves. From this circumstance we infer that the two curves actually have  equal shapes
and that they can be made to overlap by a translation $\Delta T$ along the temperature axis. From the cross-correlation plot, one can estimate  $\Delta T=n_{\rm max} \times s_T$,  where $n_{\rm max}$ is the shift
for which the cross-correlation reaches its maximum.

\subsection{Grid sizing and error estimates}
\label{stripes}

\begin{figure}
\includegraphics[width=0.95\linewidth]{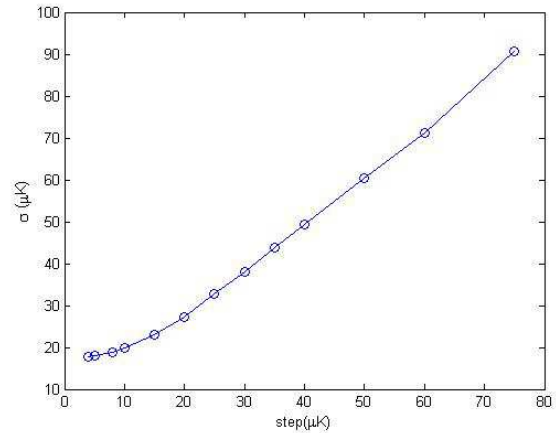}
\caption{Width $\sigma$ of the autocorrelation versus the grid step $s_T$.}
\label{step}
\end{figure}

To choose the best size of the grid cells,  we investigated the behavior of the self-correlation on a trial curve for several different choices
of the steps $s_T$ along the temperature axis.  We pointed out earlier that the distribution of bins around
the autocorrelation peak is Gaussian shaped. We found that the  width  $ \sigma$ of the
Gaussian decreases with the grid step until it stabilizes, as shown in Fig. \ref{step}.
On the basis of this behavior, a step of $15 \mu K$  in the plateau region was chosen, which
is still large enough to ensure that the typical
number of points in the occupied cells of the grid is  five or six.

The error on the  temperature shifts $\Delta T$ between any two curves $R(T)$ was estimated by using the familiar jackknife method.
We divided the transition region of the $T-R$ plane into $n$ stripes parallel to the $T$-axis, and of equal widths along the $R$-axis, and we covered
each of these stripes with  grids of equal steps $s_T$ and $s_R$. The data points falling in  each of the $n$ stripes were then analyzed by the cross-correlation method,
providing $n$ independent estimates $\Delta T_k^{(n)}$, $k=1,\dots,n$ of the temperature shifts.  For each number of subdivisions $n$, we then determined the corresponding average  shift $\Delta T^{(n)}
=\sum_{k=1}^n \frac{\Delta T_k^{(n)}}{n}$, and  the variance
$\sigma_{{\Delta T}}^{(n)}=\sqrt{\frac{\sum_{k=1}^{n}
{({\Delta T}_k^{(n)}-\Delta T^{(n)})^2}}{n-1}}$.
To find the optimal number of stripes,  the above process was repeated for  \textit{n}$\in\{1,40\}$.
In Fig. \ref{stripes} we plot the behavior of $\sigma_{\Delta T}^{(n)}$ versus \textit{n}. As we see,
$\sigma_{\Delta T}^{(n)}$ reaches a plateau for $n\approx10-15$, and on this basis we adopted $n=15$ for our
final assessment of the errors.
\begin{figure}
\includegraphics[width=0.95\linewidth]{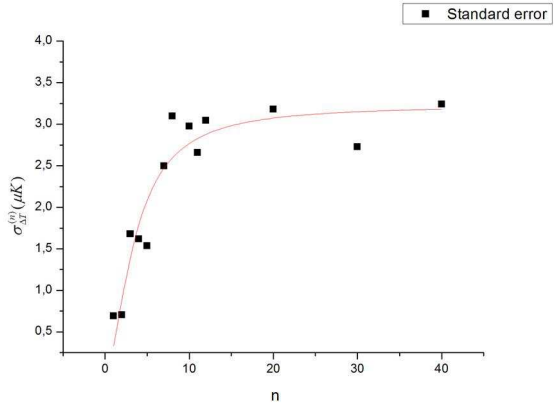}
\caption{The error $\sigma_{\Delta T}^{(n)}$ versus the number $n$ of subdivisions. The error reaches a plateau
for $n\approx$ 10-15.}
\label{stripes}
\end{figure}

Having determined the optimal grid size and number of stripes, we could then determine the best estimates for the
differences $\Delta T_c=T_c(H)-T_{c\,0}$ between the critical temperatures in the presence of the magnetic field and in null field, with the
relative errors.   For each triplet of measurements as described in Sec. III (see Fig.   \ref{R(T)measur}), we estimated
\begin{equation}
\Delta T_c= \Delta T_{(1,2)}-\frac{\Delta T_{(1,3)}}{2}\;,
\label{eq:deltaT}
\end{equation}
where the superscripts 1 and 3 refer to the curves in zero field, and the superscript 2 to the curve in the presence of the applied field. The error on $\Delta T_c$ was taken to be
\begin{equation}
\sigma_{\Delta T_c}=\sqrt{\sigma_{\Delta T_{(1,2)}}^2
+\left(\frac{1}{2}\sigma_{\Delta T_{(1,3)}}\right)^2
-{\rm cov}_{(\Delta T_{(1,2)},\Delta T_{(1,3)})}}
\label{eq:sigma}
\end{equation}
where $\sigma_{\Delta T_{1,2}}$ and $\sigma_{\Delta T_{1,3}}$ are the
standard deviations obtained as explained previously with $n=15$.

\section{Experimental results}
\label{exp_results}

\begin{figure}
\includegraphics[width=0.95\linewidth]{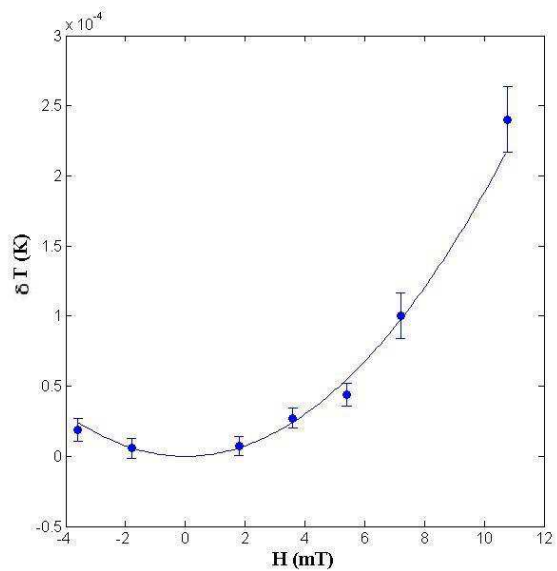}
\caption{Bare film data: the points follow the expected parabolic behavior.}
\label{film}
\end{figure}

The first thing that we checked is that the data for the bare film actually follow
the theoretical law Eq. (\ref{H_par}). According to that relation, the shifts $\delta T$ should have a parabolic dependence
on the applied magnetic
field $H$. This expectation is fully verified by our data, as  can be seen from
Fig.   \ref{film}, where the data are plotted together with a parabolic fit (continuous line).

\begin{figure}[!h]
\includegraphics[width=0.95\linewidth]{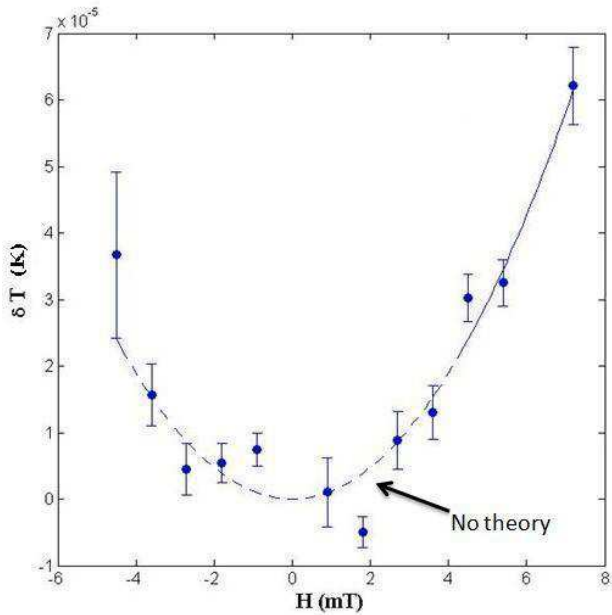}
\caption{In-cavity data together with a parabolic fit on
higher magnetic field data: for low magnetic field, corresponding to 
the dashed part of the curve, the data do not follow the parabolic behavior.}
\label{cavity}
\end{figure}

The data for the in-cavity film are shown in Fig. \ref{cavity} (note the different scales for $H$ and $\delta T$ in comparison with Fig. \ref{film}) together with a parabolic fit on
higher magnetic field data.  It is apparent that   low-field in-cavity data show deviations from the parabolic behavior, unlike the bare-film data.

\begin{figure}[!h]
\includegraphics[width=0.95\linewidth]{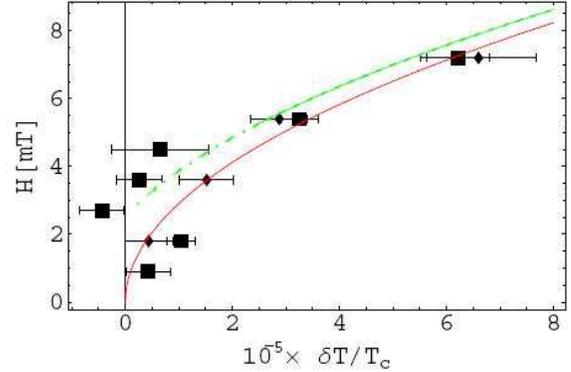}
\caption{Theoretical prediction and experimental results. In-cavity film
data (squares), bare film data (diamonds). The lower curve shows the
theoretical prediction for bare film data, the upper one that for in-cavity
film data. The point-dashed line indicates the region where a definite
theoretical prediction is not possible.}
\label{results}
\end{figure}

The different behavior of
in-cavity data compared to bare-film data can be better appreciated from Fig. \ref{results}, where they are both plotted.  It should be noted that in Fig. \ref{results} the shifts $\delta T/T_c$ are reported as a function of the
absolute value of the magnetic field, the shifts being independent of the sign of $H$. The two curves in Fig. \ref{results} are the same as in Fig. \ref{expected}. We observe again that the bare-film data lie nicely on the expected theoretical parabolic curve.
More detailed comments  are  in order for the  in-cavity  data in Fig.
\ref{results}. Let us consider first the region corresponding to
$H\approx 5\div 6$ mT: as discussed in Sec. II, for these larger fields  the Casimir energy variation is
sufficiently small compared to the condensation energy to  justify  our perturbative calculations.  In this  region one expects  a small deviation of the
in-cavity data from the bare-film parabolic behavior.
 Unfortunately,  the error bars
are of the same order of magnitude of the expected small deviations, and therefore
a  better sensitivity would be needed to ascertain the effect in this region.
For lower magnetic fields, the condensation energy and the variation of Casimir energy
become of the same order, the perturbation expansion is  no longer possible,
and deviations are possibly larger.
Interestingly, in this energy region, the data of the in-cavity film are no
longer compatible with a parabolic behavior, as   is the case for the bare film.
This suggests the hypothesis that in-cavity data display an anomalous trend
with respect to bare film data. Note that the preliminary analysis reported
in \cite{low_noise}, even if less rigorous, is compatible with the present
results. However, it is not possible to draw final conclusions from this
observation, since error bars are too  large and a theoretical model for low
condensation energy is   missing.

To sum up, in-cavity data show a behavior which is qualitatively different from the
bare film case, and   compatible with the predictions of Refs.
\cite{variation,towards}.  Even if we cannot draw  definite conclusions from
the present analysis, its results encourage us in  continuing this research,
and exploring different configurations to enhance the effect.

\section*{Acknowledgments}
We would like to thank Professor F. Gatti and the INFN Genova group for
collaboration in making the samples of Tungsten and Iridium which will be
used to realize the upgrade of the experiment. G. Esposito is grateful to
Dipartimento di Scienze Fisiche of Federico II University for hospitality
and support.


\end{document}